# Thermal stability and phase transformation of α-, κ(ε)-, and γ-$Ga_2O_3$ thin films to β-$Ga_2O_3$ under various ambient conditions

Jingyu Tang,[1,a)] Kunyao Jiang,[1] Po-Sen Tseng,[1] Rachel C. Kurchin,[1] Lisa M. Porter,[1] and Robert F. Davis[1,2]

**Affiliations**

[1]Department of Materials Science and Engineering, Carnegie Mellon University, Pittsburgh, Pennsylvania 15213, USA

[2]Department of Electrical and Computer Engineering, Carnegie Mellon University, Pittsburgh, Pennsylvania 15213, USA

[a)]Author to whom correspondence should be addressed: jingyuta@andrew.cmu.edu

**Abstract**

Phase transitions in metastable α-, κ(ε)-, and γ-$Ga_2O_3$ films to thermodynamically stable β-$Ga_2O_3$ during annealing in air, $N_2$, and vacuum have been systematically investigated via in-situ high-temperature X-ray diffraction (HT-XRD) and scanning electron microscopy (SEM). These respective polymorphs exhibited thermal stability to ~471-525°C, ~773-825°C, and ~490-575°C before transforming into β-$Ga_2O_3$, across all tested ambient conditions. Particular crystallographic orientation relationships were observed before and after the phase transitions, i.e. (0006) α-$Ga_2O_3$ → ($\bar{4}$02) β-$Ga_2O_3$, (004) κ(ε)-$Ga_2O_3$ → (310) and ($\bar{4}$02) β-$Ga_2O_3$, and (400) γ-$Ga_2O_3$ → (400) β-$Ga_2O_3$. The phase transition of α-$Ga_2O_3$ to β-$Ga_2O_3$ resulted in catastrophic damage to the film and upheaval of the surface. The respective primary and possibly secondary causes of this damage are the +8.6% volume expansion and the dual displacive and reconstructive transformations that occur during this transition. The κ(ε)- and γ-$Ga_2O_3$ films converted to β-$Ga_2O_3$ via singular reconstructive transformations with small changes in volume and unchanged surface microstructures.

Gallium oxide ($Ga_2O_3$) occurs in four universally accepted polymorphs: α-(trigonal), β-(monoclinic), γ-(cubic-defective spinel), and κ(ε)-(orthorhombic) $Ga_2O_3$. The existence of a fifth polymorph, δ-$Ga_2O_3$, with a cubic-bixbyite structure remains under debate.[1] Among these polymorphs, β-$Ga_2O_3$ is the thermodynamically stable phase to its melting point; all other polymorphs convert to β-$Ga_2O_3$ when heated.[2] While β-$Ga_2O_3$ has attracted the most attention for its potential in high-power electronics and optoelectronics applications, the monoclinic symmetry of this polymorph creates anisotropy in its optical, electronic, and thermal properties.[3–6] As such, the metastable polymorphs with their distinct crystal structures, symmetries, and properties have recently garnered considerable interest.

α-$Ga_2O_3$ has the largest bandgap (~5.3 eV[7,8]) among all the polymorphs and is a promising alternative material for applications in high-power electronics, as indicated by its higher Baliga figure of merit relative to β-$Ga_2O_3$.[9] Electrical measurements of Schottky barrier diodes fabricated on α-$Ga_2O_3$ have shown a lower on-resistance of ~0.1 mΩ·$cm^2$ (Ref. 10) compared with >2 mΩ·$cm^2$ for β-$Ga_2O_3$.[11,12] α-$In_2O_3$/α-$Ga_2O_3$ p-n junctions have also been reported.[13,14] Moreover, the bandgap of α-$Ga_2O_3$ can be tuned via alloying with other oxides such as $Al_2O_3$ and $In_2O_3$[9,15] that possess the same corundum structure. κ(ε)-$Ga_2O_3$ is also of considerable interest as it is the only polymorph that lacks inversion symmetry, meaning that spontaneous polarization is theoretically possible and this material could be piezoelectric (and possibly ferroelectric).[16,17] If ferroelectricity were to occur in κ(ε)-$Ga_2O_3$, the magnitude has been predicted to be an order of magnitude higher than in AlN and GaN.[18,19] Prior work on γ-$Ga_2O_3$ has indicated its potential utility in applications such as photoluminescence, catalytic and photocatalytic degradation of select organic compounds, and room-temperature ferromagnetism when incorporated with Mn.[20–22] For these and other applications, it is crucial to understand maximum operating temperatures for these metastable polymorphs, as well as the upper limits of the post-fabrication thermal treatments such as rapid thermal annealing of metal contacts and post-annealing after ion implantation for dopant activation.

Several studies have examined the thermal stability of α- and κ(ε)-$Ga_2O_3$ across varying film thicknesses and annealing processes, including both ex-situ and in-situ methods along with different ramping/cooling rates and holding periods at specific temperatures.[23–27] In this Letter, a comprehensive analysis of the thermal stability of α-, κ(ε)-, and γ-$Ga_2O_3$ epitaxial films, maintained at a similar thickness and subjected to an identical in-situ annealing sequence, was undertaken under three ambient conditions. The effect of the annealing environment on thermal stability and the evolution of surface morphologies are detailed for each polymorph.

The epitaxial films investigated in this Letter were α-$Ga_2O_3$ grown on (0001) sapphire via halide vapor phase epitaxy (HVPE) by Kyma Technologies, and κ(ε)- and γ-$Ga_2O_3$ grown on (0001) sapphire and (100) $MgAl_2O_4$ (spinel) substrates, respectively, via vertical, low-pressure, cold-wall metal-organic chemical vapor deposition (MOCVD). The details regarding the precursors and parameters employed for the growth of our κ(ε)- and γ-$Ga_2O_3$ films have been reported in previous publications.[28,29] High-resolution X-ray diffraction (HR-XRD) patterns of the as-grown films were acquired using a Panalytical X'pert Pro MPD X-ray diffractometer having a 4-crystal Ge×(220) monochromator. HT-XRD patterns were recorded with a Malvern Panalytical Empyrean diffractometer (mirror optics) equipped with a HTK 16N high-temperature chamber that can be heated to 1600°C. The latter equipment was employed to observe phase transitions in our metastable polymorphs to β-$Ga_2O_3$ as a function of temperature in air, ultra-high purity (5N) $N_2$ and vacuum (~$5\times10^{-5}$ Torr). Each $Ga_2O_3$ film/substrate assembly was heated in increments of 100°C and held at each temperature for 1 hr while the XRD data were acquired. Heating in increments of 25°C was applied near the phase transition temperature to observe the dynamics of this transition in each metastable polymorph. The ramping rate was set at 30°C/min to 500°C and reduced to 10°C/min for higher temperatures. The cooling rate was set to be 30°C/min. Before initiating in-situ measurements in $N_2$, the HT-XRD chamber underwent five pump-and-purge cycles to ensure the elimination of residual gases. As discussed below, temperature calibration under vacuum was necessary after the measurement due to the changes in the modes of heating in this environment. Scanning electron microscopy (SEM) was performed to investigate the evolution of the surface morphology of the as-grown and post-annealed films for each polymorph.

Figures 1(a)-1(c) show 2θ-ω scans of α-$Ga_2O_3$, κ(ε)-$Ga_2O_3$, and γ-$Ga_2O_3$ films grown on c-plane sapphire, c-plane sapphire, and (100) $MgAl_2O_4$, respectively. Figure 1(b) reveals a shoulder near the (004) κ(ε)-$Ga_2O_3$ peak, attributed to the presence of $(\bar{4}02)$ β-$Ga_2O_3$ within the film, a common observation for κ(ε)-$Ga_2O_3$ films grown by MOCVD.[28,30,31] β-$Ga_2O_3$ has been shown to be present near the film-substrate interface.[28,31] Jinno et al.[24] reported a strong correlation between film thickness and thermal stability of α-$Ga_2O_3$ films deposited on c-plane sapphire; the thinner the film, the more stable upon heat treatment. To mitigate thickness and substrate-induced strain effects on thermal stability of our films, the thicknesses were maintained at ~0.9μm and confirmed to be in a relaxed state in reciprocal space maps (RSMs) shown in Figs. 1(d)-1(f). The separation of the vertical line passing through the center of the film spot (black dashed line) and substrate spot (red dashed line) indicates that these films were in a relaxed state.

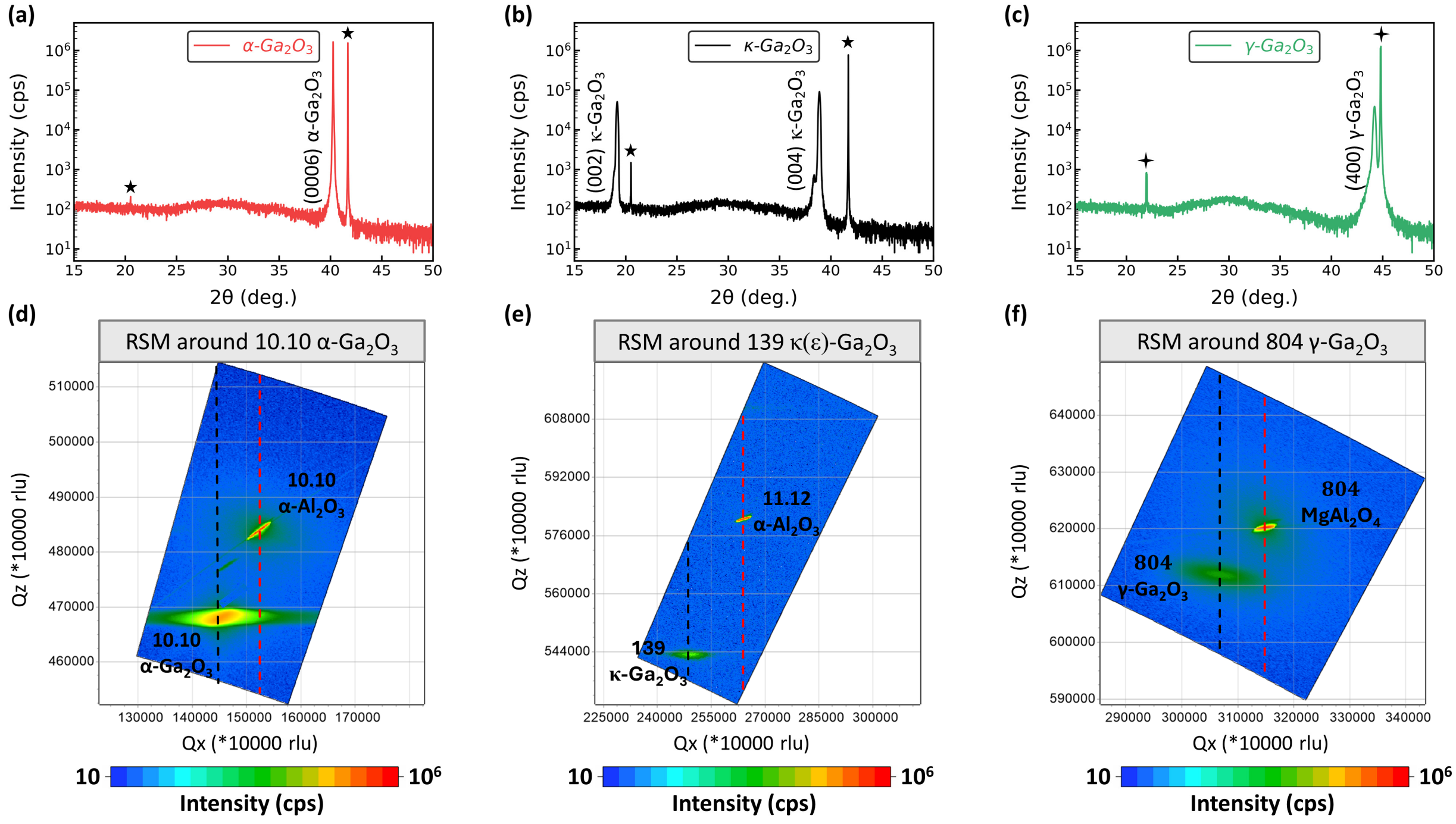

Figure 1: High-resolution 2θ-ω scans displayed on a log scale for (a) α-$Ga_2O_3$ on c-sapphire, (b) κ(ε)-$Ga_2O_3$ on c-sapphire, and (c) γ-$Ga_2O_3$ on (100) $MgAl_2O_4$ as-grown films, and their associated RSMs in (d)-(f). The starred peaks in the X-ray patterns are from the respective substrates.

Figure 2(a) shows the in-situ HT-XRD patterns of α-$Ga_2O_3$ annealed in air in the temperature range from 500°C to 900°C. These patterns reveal that (0001) oriented α-$Ga_2O_3$ maintains its thermal stability to 525°C. However, a phase transition is initiated at ~550°C, which results in the formation of ($\bar{2}01$) oriented β-$Ga_2O_3$. The results of six consecutive scans for 10min each conducted over 1 hr annealing periods at both 525°C and 550°C are illustrated in Figs. 2(a1) and 2(a2), respectively. Notably, the α-$Ga_2O_3$ film annealed at 525°C exhibits no detectable phase change, as evidenced by the nearly constant intensity of the (0006) α-$Ga_2O_3$ peak, depicted in Fig. 2(a1). In contrast, Figure 2(a2) reveals a noticeable decrease in the intensity of the (0006) α-$Ga_2O_3$ peak accompanied by a gradual increase in the intensity of the ($\bar{4}02$) β-$Ga_2O_3$ peak for the film annealed at 550°C. The α-$Ga_2O_3$ film is completely converted to β-$Ga_2O_3$ after annealing at 600°C, because the α-$Ga_2O_3$ peak is no longer detectable and further increases in temperature do not affect the peak intensity of ($\bar{4}02$) β-$Ga_2O_3$. Wen et al.[23] determined using in-situ HT-XRD that a 4μm thick α-$Ga_2O_3$ grown on c-plane sapphire by HVPE was thermally stable up to 500-540 °C in air. Studies by other investigators[24,25] also demonstrated enhanced thermal stability of α-$Ga_2O_3$ beyond 600°C via reduction of the film thickness to ~100 nm on c- and m-plane sapphire. Furthermore, they determined that either the application of the selective area growth method to release the strain within the film[24] or the deposition of an $AlO_x$ capping layer to suppress the decomposition in tandem with the application of strain to the film[25] further extended the thermal stability of α-$Ga_2O_3$ to within the temperature range of 800-900°C.

Figure 2(b) presents the in-situ HT-XRD patterns of κ(ε)-$Ga_2O_3$ annealed in air in the same temperature range from 500°C to 900°C. The (001) oriented κ(ε)-$Ga_2O_3$ was thermally stable to 800°C, with a phase transition to both $(\bar{4}02)$ and (310) oriented β-$Ga_2O_3$ starting around 825°C. This transition is detailed in the six successive scans over a 1 hr annealing period at both 825°C and 850°C, as shown in Figs. 2(b1) and 2(b2), respectively. Annealing at 825°C resulted in a gradual decrease in the intensity of the (004) κ(ε)-$Ga_2O_3$ peak and an increase in the intensity of both $(\bar{4}02)$ and (310) β-$Ga_2O_3$ peaks towards the end of the 1 hr annealing period, as shown in Fig. 2(b1). The transition occurred more rapidly for the film annealed at 850°C (Fig. 2(b2)). κ(ε)-$Ga_2O_3$ was completely converted to β-$Ga_2O_3$, as there was no detectable κ(ε)-$Ga_2O_3$ peak and no discernible increase in the peak intensity of either the $(\bar{4}02)$ or the (310) β-$Ga_2O_3$ peaks when the temperature exceeded 850°C. The coexistence of $(\bar{4}02)$ and (310) β-$Ga_2O_3$ may be attributed to their similar oxygen sublattice atomic arrangements and lattice mismatches with (0001) α-$Ga_2O_3$. Studies by Fornari et al.[26] revealed that cooling κ(ε)-$Ga_2O_3$ films from 1000°C at a rate of 7.5°C/min and 2°C/min resulted in the evolution of either the (310) or $(\bar{4}02)$ peak, respectively, as the dominant β-$Ga_2O_3$ peak. Recently, Xu et al.[32] explored the competitive growth of the (310) and the $(\bar{4}02)$ orientations of β-$Ga_2O_3$ grown on c-plane sapphire by HVPE. They noted the similarity of the atomic arrangements of these two planes and the c-plane of sapphire, and the effect of growth temperatures and HCl flow rates on the formation of the energetically unfavorable (310) oriented β-$Ga_2O_3$.

Figure 2(c) presents the in-situ HT-XRD patterns of γ-$Ga_2O_3$ annealed in air within the aforementioned temperature range. The (100) oriented γ-$Ga_2O_3$ was determined to be thermally stable to 575°C; however, a phase transition to (100) oriented β-$Ga_2O_3$ was initiated at ~600 °C. This transition progressed more slowly than those observed for α- and κ(ε)-$Ga_2O_3$. Successive scans over 1 hr annealing periods at 575°C and 900°C are shown in Figs. 2(c1) and 2(c2), respectively. The scans in Figure 2(c1) exhibit no significant phase change, as evidenced by the nearly constant intensity of the (400) γ-$Ga_2O_3$ peak. The small lattice mismatch (~+1.9%) between the γ-$Ga_2O_3$ film and the $MgAl_2O_4$ substrate, combined with the resolution limits of the XRD equipment, made difficult the observation of the (400) γ-$Ga_2O_3$ peak below 3600 cps, since the background intensity at the same 2θ position for a bare $MgAl_2O_4$ substrate is approximately 3600 cps. Thus, the continued increase of the (400) β-$Ga_2O_3$ peak for the film annealed at 900°C suggests an incomplete phase transition, as shown in Fig. 2(c2). Further increasing the annealing temperature to 1000°C caused significant Mg and Al interatomic diffusion from the substrate into the film and resulted in the formation of γ-$Ga_2O_3$ solid solutions, as detailed in our earlier publications.[33,34]

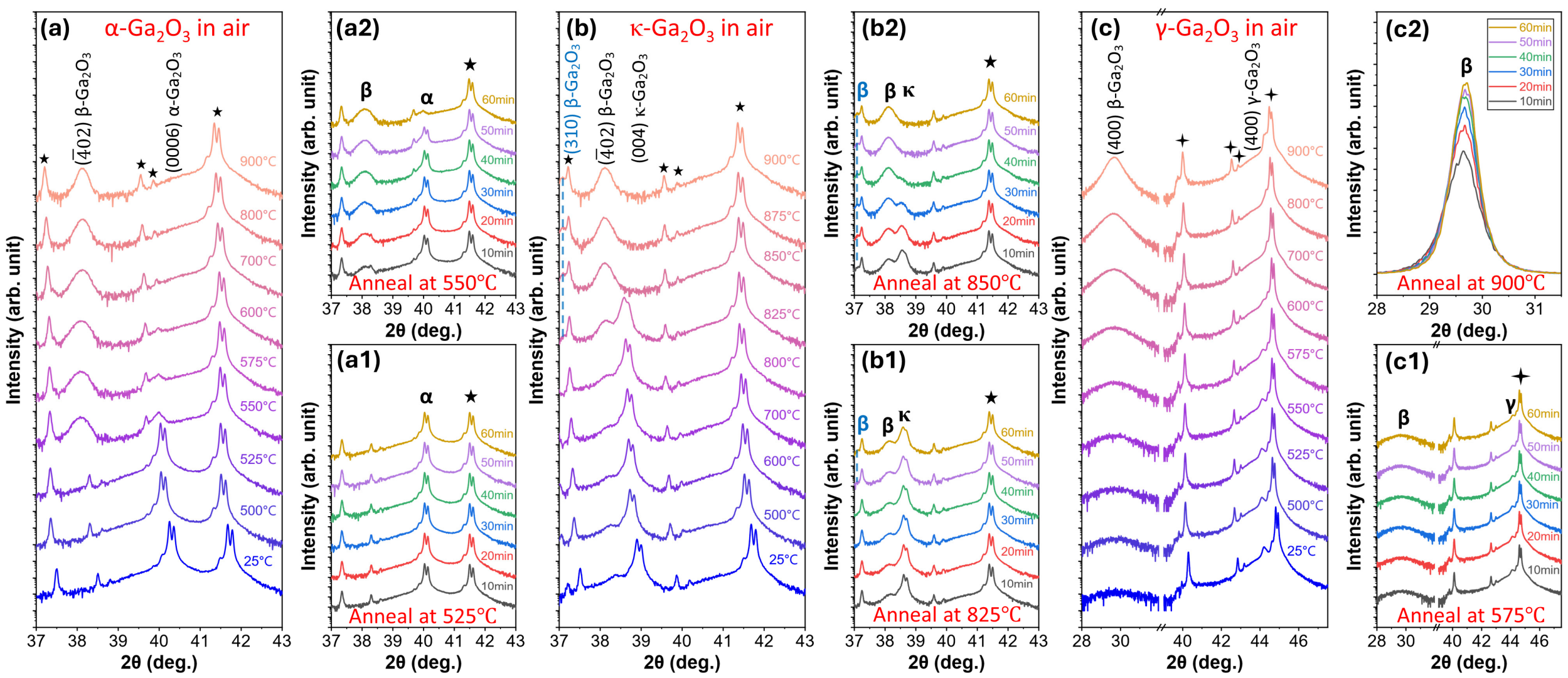

Figure 2: HT-XRD 2θ-ω scans conducted in air for (a) α-$Ga_2O_3$ on c-sapphire, (b) κ(ε)-$Ga_2O_3$ on c-sapphire, and (c) γ-$Ga_2O_3$ on (100) $MgAl_2O_4$, respectively, in the temperature range of 500-900°C. The starred peaks in the X-ray patterns are from the respective substrates.

Figures 3(a)-3(c) show the in-situ HT-XRD patterns of α-$Ga_2O_3$, κ(ε)-$Ga_2O_3$ and γ-$Ga_2O_3$ films, respectively, annealed in $N_2$ within the temperature range of 500°C to 900°C. Similar to the observations made in air, the α-$Ga_2O_3$, κ(ε)-$Ga_2O_3$ and γ-$Ga_2O_3$ films were thermally stable to approximately 525°C, 825°C, and 575°C, respectively. Almost no detectable decrease in the film peak intensities occurred during the 1 hr annealing period at these temperatures, as shown in Figs. 3(a1), 3(b1) and 3(c1). Lee et al.[27] reported that an α-$Ga_2O_3$ film with a thickness of ~720nm was thermally stable to 550°C in $N_2$. Fornari et al.[26] determined via ex-situ annealing that phase transitions in 500nm thick κ(ε)-$Ga_2O_3$ films occurred around 900°C, in both $O_2$ and $N_2$ atmospheres.

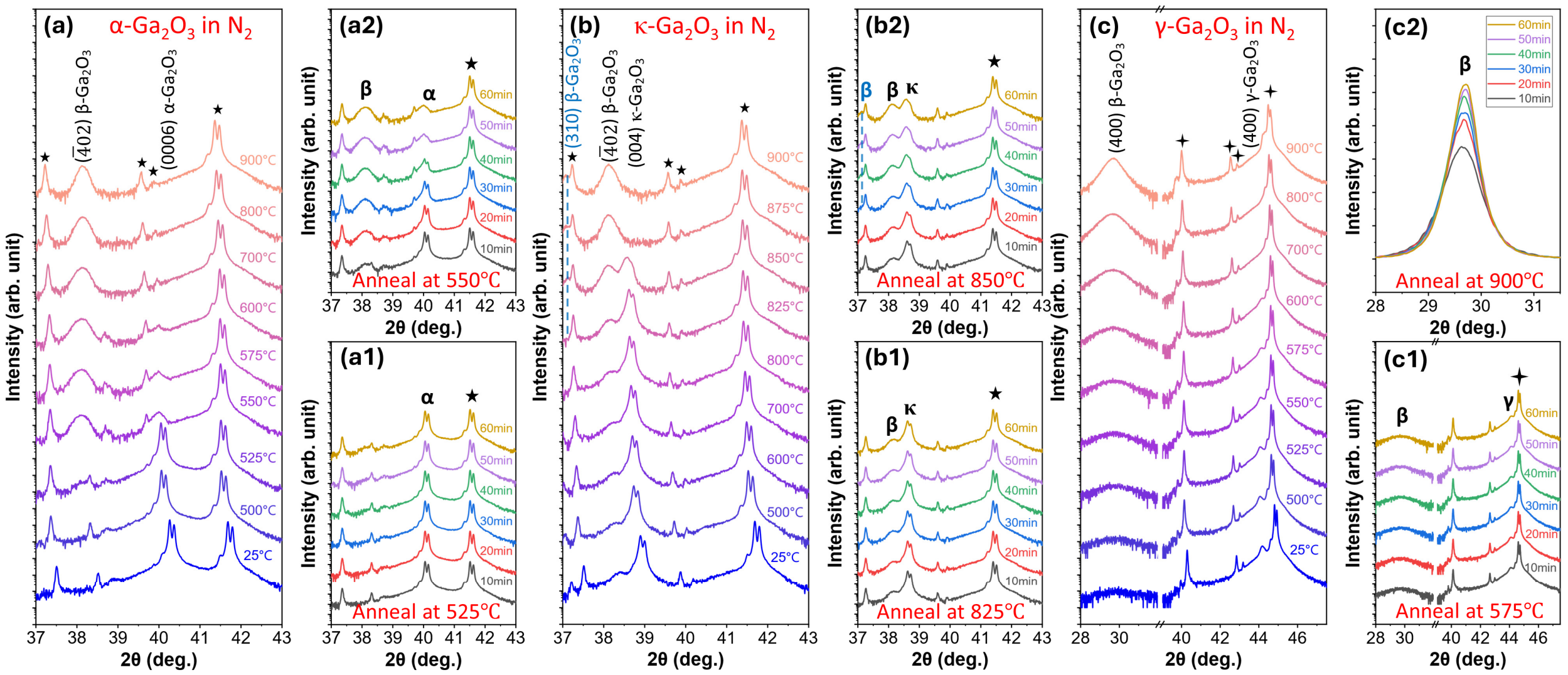

Figure 3: HT-XRD 2θ-ω scans conducted in $N_2$ for (a) α-$Ga_2O_3$ on c-sapphire, (b) κ(ε)-$Ga_2O_3$ on c-sapphire, and (c) γ-$Ga_2O_3$ on (100) $MgAl_2O_4$, respectively, in the temperature range of 500-900°C. The starred peaks in the X-ray patterns are from the respective substrates.

The in-situ HT-XRD experiments under vacuum were conducted with the sample temperature range from 110°C to 865°C based on the calibration process described below. The temperature difference between the platinum stage and the surface of the sapphire and spinel substrates was determined to be less than 20°C when the stage temperature was below 900°C in air and $N_2$. However, in conducting the same in-situ measurements in vacuum, significant temperature differences were noted between the sample stage and the substrate because of the poor heat transfer in vacuum, likely due to lack of convection to the sample surface from the atmosphere. This observation was initially identified from the inconsistency in the calculated out-of-plane lattice constants of the sapphire and spinel substrates beneath the α-$Ga_2O_3$, κ(ε)-$Ga_2O_3$, and γ-$Ga_2O_3$ films, in comparison to similar data acquired for these substrates heated in air and $N_2$, as shown in Figs. 4(a)-4(c). Temperature calibration in vacuum was done based on the second-order polynomial fit of the out-of-plane lattice vs. temperature determined in air and $N_2$. Figures 4(d)-4(f) show a consistent out-of-plane lattice constant of the sapphire and spinel substrates versus calibrated temperature.

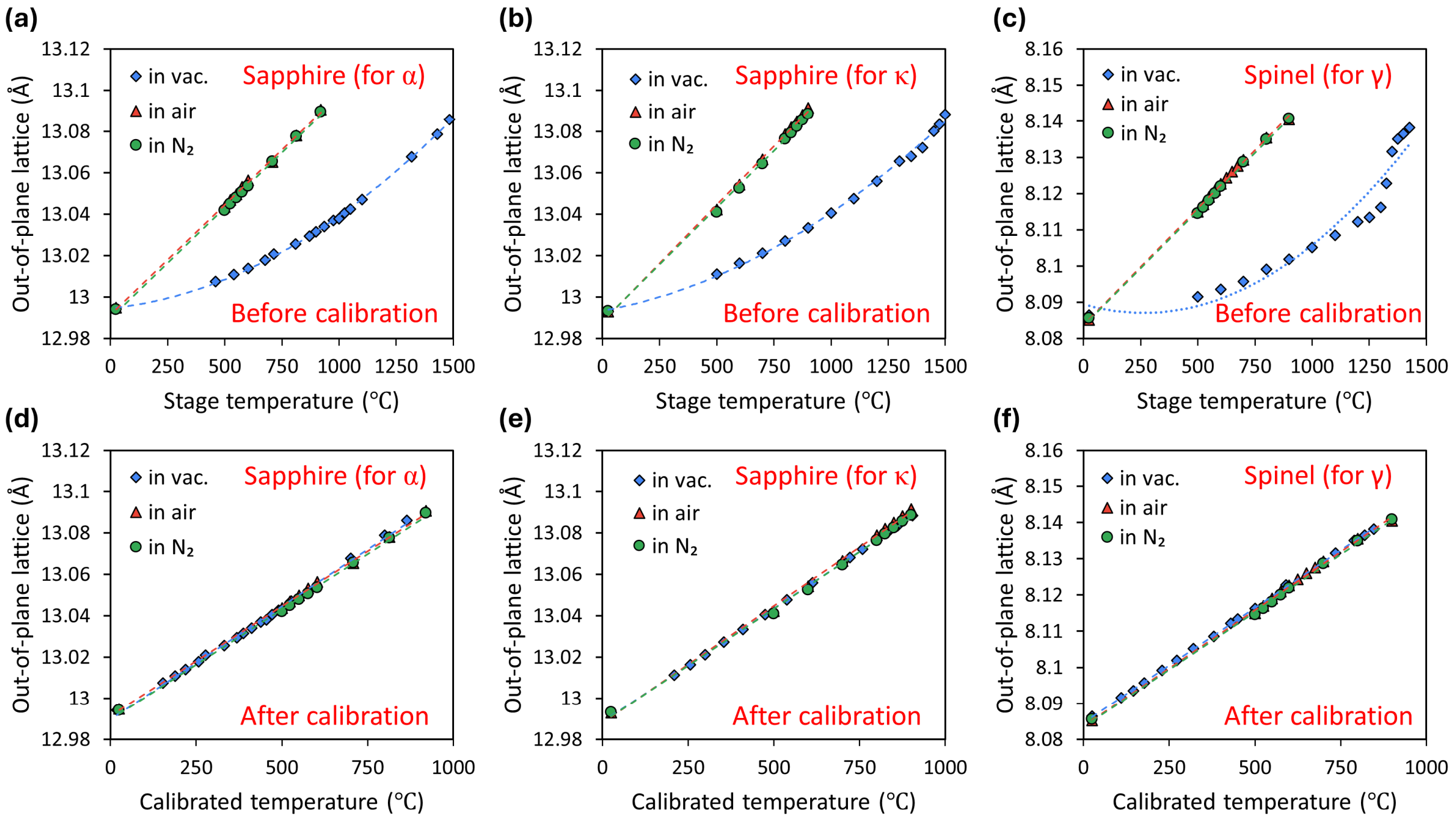

Figure 4: Out-of-plane lattice constants versus stage or calibrated temperature for sapphire and spinel substrates in vacuum before and after calibration.

Figures 5(a)-5(c) present the in-situ HT-XRD patterns of α-$Ga_2O_3$, κ(ε)-$Ga_2O_3$, and γ-$Ga_2O_3$ films annealed in vacuum from room temperature to ~900°C. Consistent with findings in air and $N_2$, the thermal stabilities of α-$Ga_2O_3$, κ(ε)-$Ga_2O_3$ and γ-$Ga_2O_3$ films were maintained within the ranges of 471-490°C, 773-818°C, and 490-590°C, respectively, as shown in Figs. 5(a1), 5(b1) and 5(c1). As noted above, Wen et al.[23] conducted a similar HT-XRD study on α-$Ga_2O_3$ films grown

on c-sapphire substrates and annealed within the temperature range of under a vacuum of $10^{-4}$ Torr. They observed that the phase transition to β-$Ga_2O_3$ occurred at 700°C, a temperature 200°C higher than annealing in air. This variation is thought to be attributed to the difference in oxygen partial pressures affecting oxygen diffusion. They surmised that the oxygen concentrations of the various intermediate $GaO_x$ compositions decrease during the transformation with the decrease in $pO_2$ due to the reduction of oxygen diffusion. This phenomenon eventually increases the overall kinetic barrier for phase transformation. However, our results did not indicate a similar enhancement in thermal stability under vacuum in any of the three metastable polymorphs. We hypothesize that this may be related to the length of time the samples were held at each temperature. Wen et al. stabilized the sample for 5 min at each temperature before recording the data; we held the sample for 1 hr, during which six successive scans were performed to verify whether the system had reached equilibrium.

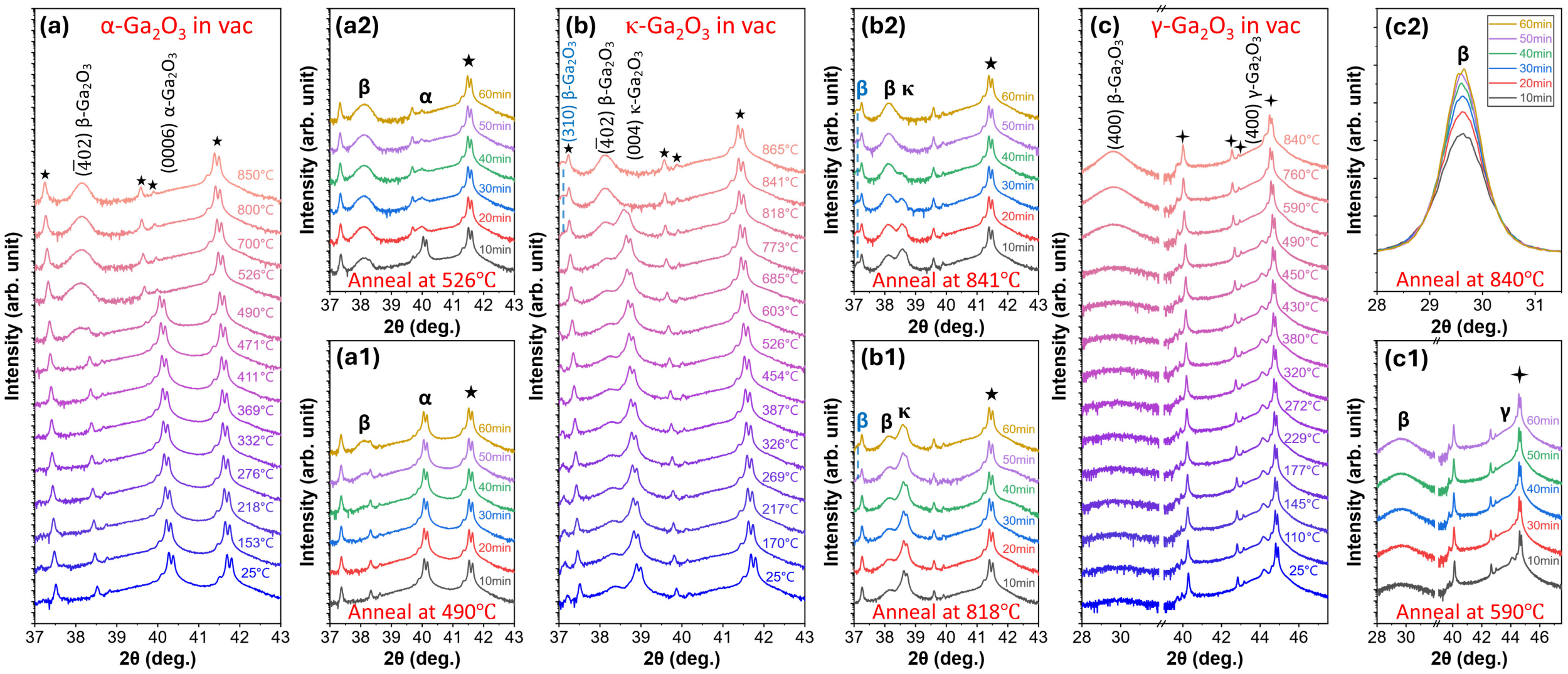

Figure 5: HT-XRD 2θ-ω scans conducted in vacuum with a base pressure of $5\times10^{-5}$ Torr for (a) α-$Ga_2O_3$ on c-sapphire, (b) κ(ε)-$Ga_2O_3$ on c-sapphire, and (c) γ-$Ga_2O_3$ on (100) $MgAl_2O_4$, respectively, in the calibrated temperature range of 110-865°C. The starred peaks in the X-ray patterns are from the respective substrates.

Figure 6 displays the SEM images of both as-grown and post-annealed α-$Ga_2O_3$, κ(ε)-$Ga_2O_3$, and γ-$Ga_2O_3$ films processed under various conditions. As detailed above, annealing the films resulted in their complete phase transformation into β-$Ga_2O_3$. Only the α-$Ga_2O_3$ films experienced a catastrophic phase transition characterized by widespread rupture and upheaval of the surface across the entire material regardless of the ambient conditions. A similar observation has been reported in α-$Ga_2O_3$ films by Wen et al.[23] This phenomenon of marked film damage is similar to that observed during the θ-alumina (C2/m, structurally equivalent to β-$Ga_2O_3$) to α-alumina ($R\bar{3}c$, structurally equivalent to α-$Ga_2O_3$) phase transition. This latter transition has been attributed to a substantial volume reduction of ~10 vol%[35,36] during heating. In the case of the α- to β-$Ga_2O_3$ transition, a positive volume expansion of +8.6 vol% is involved and is the primary cause of the severe film damage. In contrast, the κ(ε)-$Ga_2O_3$ and γ-$Ga_2O_3$ films maintained their surface morphologies after each phase transition in the three ambient conditions. The data presented in Table I provide a comparison of the change in volume of each metastable $Ga_2O_3$ phase in relation to β-$Ga_2O_3$ as well as the type of phase transition, density, and the oxygen sublattice arrangement in each phase. The transformation of α-$Ga_2O_3$ to β-$Ga_2O_3$ is both a displacive and a reconstructive phase transition, which is likely a secondary cause of the disruption in the microstructure. The shearing of the oxygen sublattice from hexagonal close-packing (HCP) to face-centered cubic (FCC) packing is classified as a diffusionless displacive (martensitic) transition.[37] However, the relocation of gallium cations involves short-range diffusion and synchronized shear with the oxygen sublattice and is classified as a reconstructive transition.[37] By comparison, the oxygen sublattices for κ(ε)-$Ga_2O_3$ and γ-$Ga_2O_3$ are FCC structures, which are structurally similar to the oxygen sublattice in β-$Ga_2O_3$. The phase transitions in these two polymorphs primarily involve short-range diffusion and are thus reconstructive transitions. Theoretical studies on the transition from γ-$Al_2O_3$ ($Fd\bar{3}m$, structurally equivalent to γ-$Ga_2O_3$) to θ-alumina suggest it involves local migration of Al atoms without the shearing of the oxygen sublattice.[38] The volume changes associated with the phase transitions in κ(ε)-$Ga_2O_3$ and γ-$Ga_2O_3$ are much smaller than that in α-$Ga_2O_3$, as shown in Table I, and contribute markedly to the preservation of the surface morphology during the phase transition in each polymorph.

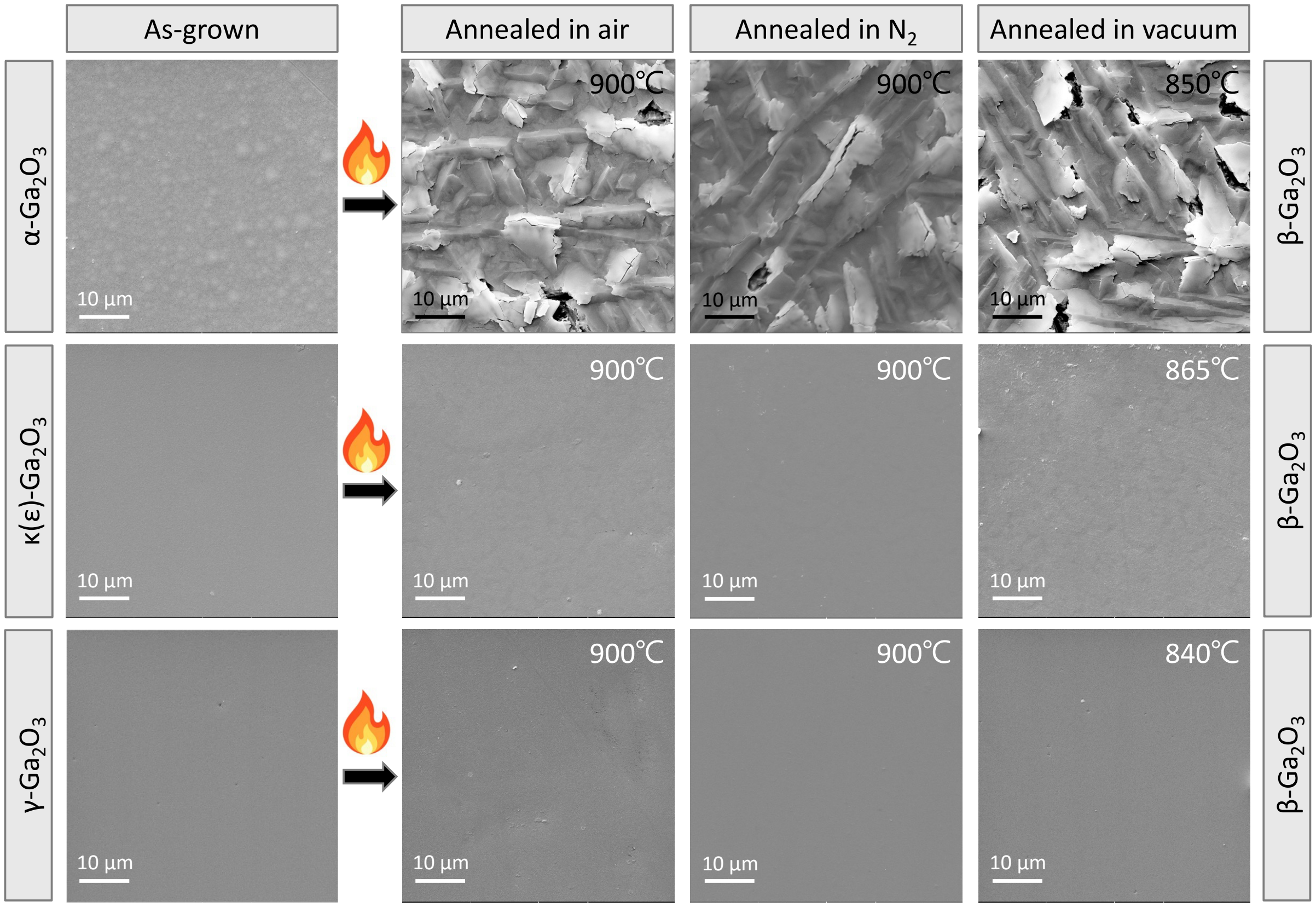

Figure 6: SEM images of the α-$Ga_2O_3$, κ(ε)-$Ga_2O_3$, and γ-$Ga_2O_3$ films before and after annealing under different ambient conditions. All films converted to β-$Ga_2O_3$ by ~900°C.

Table I. Comparison of the density, change in volume at room temperature, and type of phase transition involved in the transformation of $\alpha$-$Ga_2O_3$, $\kappa(\varepsilon)$-$Ga_2O_3$, and $\gamma$-$Ga_2O_3$ films to $\beta$-$Ga_2O_3$.

| Phase | Density ($g/cm^3$) | $\Delta$ volume with respect to $\beta$ (%) | O sublattice atomic arrangement | Phase transition type |
|---|---|---|---|---|
| $\alpha$-$Ga_2O_3$ | ~6.453[39,40] | +8.6 | HCP | displacive + reconstructive[37] |
| $\kappa(\varepsilon)$-$Ga_2O_3$ | ~6.111[41,42] | +2.8 | FCC | reconstructive |
| $\gamma$-$Ga_2O_3$ | ~5.940[41,43] | -0.05 | FCC | reconstructive |
| $\beta$-$Ga_2O_3$ | ~5.943[44,45] | 0 | FCC | - |

The thermal stability and phase evolution of structurally relaxed, ~1 μm thick films of $\alpha$-$Ga_2O_3$, $\kappa(\varepsilon)$-$Ga_2O_3$, and $\gamma$-$Ga_2O_3$ have been determined via in-situ HT-XRD in air, $N_2$, and vacuum environments. These polymorphs exhibited stability against transformation to $\beta$-$Ga_2O_3$ up to ~471-525°C, ~773-825°C, and ~490-575°C, respectively, across all tested ambient conditions. Strong crystallographic orientation relationships were observed before and after the phase transitions, namely, (0006) $\alpha$-$Ga_2O_3$ → $(\bar{4}02)$ $\beta$-$Ga_2O_3$, (004) $\kappa(\varepsilon)$-$Ga_2O_3$ → (310) and $(\bar{4}02)$ $\beta$-$Ga_2O_3$, and (400) $\gamma$-$Ga_2O_3$ → (400) $\beta$-$Ga_2O_3$. Notably, the $\alpha$-$Ga_2O_3$ film experienced a catastrophic transformation characterized by severe rupture and upheaval of the surface under all ambient conditions and primarily caused by the significant positive volume expansion during the transition. The changes in the arrangement of oxygen and gallium sublattices during the $\alpha$-$Ga_2O_3$ to $\beta$-$Ga_2O_3$ transition are complex, involving both displacive and constructive transformations that likely played a secondary role in the disruption of the microstructure. They are also complex relative to the singular reconstructive transformations and much smaller volume changes that occur in the transitions of $\kappa(\varepsilon)$-$Ga_2O_3$ and $\gamma$-$Ga_2O_3$ films to $\beta$-$Ga_2O_3$.

## Acknowledgments

This material is based upon work supported by the Air Force Office of Scientific Research (Program Manager, Dr. Ali Sayir) under Award No. FA9550-21-1-0360, and by the II-VI Foundation. The use of the Materials Characterization Facility at Carnegie Mellon University was supported by Grant No. MCF-677785. J. T. and K. J. would also like to acknowledge the Neil and Jo Bushnell Fellowship in Engineering for additional support.

## Conflict of Interest

The authors have no conflicts to disclose.

**Data Availability**

The data that support the findings of this study are available from the corresponding author upon reasonable request.